\documentclass[runningheads]{llncs}

\newif\ifEditMode

 \EditModetrue

\usepackage{preamble}
\usepackage{aliases}

\usepackage{pifont}

\begin{document}
\sloppy
\title{The Writing is on the Wall: \\Analyzing the Boom of Inscriptions and its Impact on EVM-compatible Blockchains\thanks{Most of this work was performed while the authors were at Matter Labs.}}

\titlerunning{The Writing is on the Wall}


\author{Johnnatan Messias\inst{1}
\and Krzysztof Gogol\inst{2}
\and Maria Inês Silva\inst{3,5}
\and Benjamin Livshits\inst{4}}

\authorrunning{J. Messias et al.}

\institute{MPI-SWS
\and University of Zurich
\and NOVA Information Management School
\and Imperial College London
\and Matter Labs
}

\maketitle
\begin{abstract}
  
This paper examines inscription-related transactions on Ethereum and major EVM-compatible rollups, assessing their impact on scalability during transaction surges. 
Our results show that, on certain days, inscriptions accounted for nearly 90\% of transactions on Arbitrum and ZKsync Era, while 53\% on Ethereum, with 99\% of these inscriptions involving meme coin minting.
Furthermore, we show that ZKsync and Arbitrum saw lower median gas fees during these surges. ZKsync Era, a ZK-rollup, showed a greater fee reduction than the optimistic rollups studied—Arbitrum, Base, and Optimism.

\end{abstract}
%
%
%

\section{Introduction}\label{sec:introduction}

\gls{L2} scaling solutions have emerged as a response to the scalability challenges of \gls{L1} blockchains such as Bitcoin and Ethereum. These solutions aim to enhance transaction throughput and cost efficiency while inheriting the security guarantees of their underlying \gls{L1} blockchain. \glspl{L2} execute transactions off-chain and subsequently settling state transitions on an \gls{L1} blockchain.

Despite their increasing adoption, most \gls{L2} solutions had not undergone significant stress testing until late 2023, when an unprecedented spike in transaction volume occurred. This surge was largely driven by the emergence of \stress{inscriptions}, a phenomenon inspired by Bitcoin ordinals~\cite{wang2023understanding}. Ordinals, introduced in March 2023~\cite{Domo@BRC}, enable the assignment of unique identifiers to individual satoshis, effectively transforming them into \glspl{NFT}. This concept extended beyond Bitcoin, reaching Ethereum and its major \gls{L2} rollups, where inscriptions allowed users to embed arbitrary data, such as text, images, or code, directly onto the blockchain. By late 2023, inscription transactions accounted for over 80\% of all transactions on some \gls{EVM} chains~\cite{2024DuneEVMInsctiptions}, significantly influencing network performance, transaction costs, and user behavior.

This rapid increase in transaction volume led to temporary service disruptions, with certain rollups experiencing downtime of approximately 78 minutes~\cite{Haig@TheDefiant,Tom-Arbitrum@Cointelegraph}. However, during this surge, \gls{ZK} rollups---one type of \gls{L2} blockchain---offered notably lower gas costs compared to Ethereum, making them an attractive alternative for inscription trading. Furthermore, Ethereum's Dencun upgrade in March 2024 introduced \stress{blobs} as temporary data storage, reducing gas fees for rollups but raising concerns about the long-term availability of inscription-related data~\cite{park_impact_2024}. Since blobs are designed for ephemeral state verification, inscription-related data may not persist on-chain, relying instead on off-chain indexers managed by platform creators.

In this paper, we investigate the performance and user behavior of \gls{EVM}-compatible chains during the inscription boom. Our study focuses on Ethereum and its four major rollups---Arbitrum, Base, Optimism, and ZKsync Era. We address the following research questions:

\paraib{What caused the transaction surge on \gls{EVM} chains at the end of 2023?}
We analyze transaction data of Ethereum and its four major rollups. We identified that these transactions were due to inscription-based meme coins and were minted and traded at significantly lower costs compared to Bitcoin ordinals or Ethereum inscriptions. Our findings indicate that these inscriptions adopted modified versions of the BRC-20 standard~\cite{BRC-20@Binance,wang2023understanding,yu2023bridging}, facilitating their integration into NFT marketplaces~\cite{BigInt@Marketplace,Inscribe@Marketplace,ZKSMarket@Marketplace}. Specifically, on December~17,~2023, over~88\% of all transactions on ZKsync Era were due to inscriptions. Other blockchains also exhibited a high percentage of inscription-related transactions. Arbitrum recorded the highest percentage, with over~89\%, followed by Ethereum at over~53\%, Base at over~37\%, and Optimism, with over~35\% of the total inscription transactions.
    
\paraib{What impact did inscription transactions have on the cost and performance of \gls{EVM} blockchains?}
We evaluate how inscription-driven transaction spikes affected network stability, gas fees, and transaction throughput. Our analysis shows that, unlike Ethereum, where transaction surges led to increased gas fees, some rollups experienced reduced median transaction costs. We also compare the cost efficiency of ZK rollups versus optimistic rollups.
    
\paraib{Why did users choose inscription-based meme coins over ERC-20 tokens?}
We examine the user interaction with inscriptions, finding that inscription transactions predominantly involved minting rather than trading. Across all analyzed chains, over~99\% of inscription-related transactions were for minting activities, with some tokens, such as \textit{zrc-20 sync}, sustaining trading activity beyond the initial boom. For example, in comparison to the overall number of inscription transactions, the percentage of inscription minting or claiming activities was as follows: Arbitrum at~99.82\%, Base at~99.99\%, Ethereum at~99.92\%, Optimism at~99.98\%, and ZKsync Era at~99.91\%.

To promote scientific reproducibility, we make our datasets and analysis scripts publicly available~\cite{Messias-DataSet-Code-2025}. By providing insights into the inscription-driven stress test of \gls{EVM} chains, our study contributes to the broader understanding of blockchain scalability, economic incentives, and data availability in the evolving blockchain ecosystem.

\section{Background on Inscriptions}\label{sec:background_inscriptions}

Inscriptions involve the recording of arbitrary data on the blockchain. On \gls{EVM}-compatible chains, users encode \stress{HEX} data into the transaction input \stress{call data} (refer to column \stress{JSON} in Table~\ref{tab:inscriptions}), usually also setting the transaction's \stress{from} and \stress{to} attributes with the same user addresses. This structure constitutes a self-transfer made from a user to their own address. 

\subsection{Operation Types}
There are various protocol standards that define the structure of inscriptions. \stress{BRC-20} was the first standard proposed for Bitcoin, and includes three main types of operations: \stress{deploy}, \stress{mint}, and \stress{transfer}, each encoded in a single transaction~\cite{wang2023brc}. The protocol standards on \gls{EVM}-compatible chains extend this set of operations by adding, for example, \stress{claim}, \stress{list}, \stress{buy}, \stress{sell}, among others.

\begin{table}[t]
\centering
\caption{Example of inscriptions data recorded on-chain: \stress{protocol name}, \stress{operation}, \stress{token name (Tick)}, \stress{total supply}, and \stress{maximum amount of tokens minted each round (Limit)}.}
\resizebox{\textwidth}{!}{%
\begin{tabular}{cccrrl}
\toprule
\thead{Protocol} & \thead{Tick} & \thead{Operation} & \thead{Total Supply} & \thead{Limit} & \thead{JSON data}    \\
\midrule
    zrc-20 & sync  & deploy & $21\times10^6$ & \num{4} & \scriptsize{\texttt{\{"p":"zrc-20","op":"deploy","tick":"sync","amt":"21000000","limit":"4"\}}}\\
    zrc-20 & sync  & transfer & --- & \num{4} & \scriptsize{\texttt{\{"p": "zrc-20", "op": "transfer", "tick": "sync", "amt": "4"\}}}\\
    zrc-20 & sync  & mint & --- & \num{4} & \scriptsize{\texttt{\{"p":"zrc-20","op":"mint","tick":"sync","amt":"4"\}}}\\
    zrc-20 & zksi  & deploy & $21\times10^8$ & $1\times10^4$ & \scriptsize{\texttt{\{"p":"zrc-20","op":"deploy","tick":"zksi","max":"2100000000","limt":"10000"\}}}\\
    zrc-20 & zkss  & deploy & $21\times10^6$ & $1\times10^3$ & \scriptsize{\texttt{\{"p":"zrc-20","op":"deploy","tick":"zkss","max":"21000000","lim":"1000"\}}}\\
    zrc-20 & zkss  & sell & --- & \num{1000} & \scriptsize{\texttt{\{"p":"zrc-20","op":"sell","orderMessage":[\{"amt":10000,"nonce":"62...",}}\\
    zrc-20 & zkzk  & deploy & $21\times10^6$ & $21\times10^6$ & \scriptsize{\texttt{\{"p":"zks-20","op":"deploy","tick":"zkzk","max":"21000000","lim":"21000000"\}}}\\
    era-20 & bgnt  & deploy & $21\times10^6$ & \num{5} & \scriptsize{\texttt{\{"p":"era-20","op":"deploy","tick":"bgnt","max":"21000000","lim": "5"\}}}\\
    era-20 & bgnt  & deploy & $21\times10^6$ & \num{5} & \scriptsize{\texttt{\{"p":"era-20","op":"deploy","tick":"bgnt","max":"21000000","lim": "5"\}}}\\
    era-20 & bgnt  & list & --- & \num{5} & \scriptsize{\texttt{\{"p":"era-20","op":"list","tick":"bgnt","amt":"250","price":"1500000000000000"\}}}\\
    era-20 & bgnt  & buy & --- & --- & \scriptsize{\texttt{\{"p":"era-20","op":"buy","tx":"0xda..."\}}}\\
    layer-2 & \$L2  & claim & --- & \num{1000} & \scriptsize{\texttt{\{"p":"layer2-20","op":"claim","tick":"\$L2","amt":"1000"\}}}\\
\bottomrule
\end{tabular}
}
\label{tab:inscriptions}
\end{table}

\paraib{Deploy.}
The \stress{deploy} action specifies the protocol name, token tick, total supply, and the maximum amount of tokens a user can mint (or claim) per transaction. Table~\ref{tab:inscriptions} shows  examples of inscriptions' data recorded on-chain. For example, to deploy an inscription, a transaction containing a \stress{deploy} action should be recorded on-chain, marking the initiation of an inscription event. In the case of a \stress{zrc-20 sync} inscription (see first row in Table~\ref{tab:inscriptions}), the protocol specifies a total supply of~21 million inscriptions and a limit of~4 tokens minted per transaction.

\paraib{Mint.}
After the transaction that deploys the inscription persists in the chain, users can issue a \stress{mint} action to actually mint (or in this case claim ownership) of the tokens. To initiate this, users need to issue a transaction with an input call data encoded to \stress{HEX code}, specifying the protocol and tick that jointly identify the inscription-based token. In the same transaction, users provide the number of tokens they want to claim (refer to column \stress{Limit} in Table~\ref{tab:inscriptions}) that must not exceed the maximum limit specified in \stress{deploy} operation for the given protocol-tick pair. The off-chain \stress{indexer} is responsible for ensuring the integrity and user balances of inscription-based tokens.

\paraib{Claim.} Similarly to \stress{mint}, it allows users to mint and claim ownership of a new inscription token. This operation is used, for example, within the \stress{layer2-20} standard to mint a new inscription token, \stress{\$L2}. The \$L2 token is a multi-\gls{L2} inscription token that its inscription creator allows to be minted on multiple blockchains. However, there is just one \stress{deploy} operation that defines the token's maximum supply. The \stress{off-chain indexer} maintained by the creator is responsible for ensuring the integrity of data in these circumstances.

\paraib{Transfer.} Once users successfully mint the inscriptions, they can transfer ownership of their inscription tokens to another address. To do so, they typically issue a transaction to another address and add the data formatting standard of a \stress{transfer} to the input call data. The receiving address will then possess ownership of these inscriptions. The inscription token is also identified by a protocol standard and a token name (\stress{tick}). Tokens with the same names (ticks) can be deployed multiple times using various protocol standards.

\paraib{List.} It enables users to list their inscriptions on marketplaces~\cite{BigInt@Marketplace,Binance@Ordinals,Inscribe@Marketplace,ZKSMarket@Marketplace}. Users specify the quantity of tokens and their price (in Ether) that they are willing to receive in exchange for their inscriptions from interested buyers.

\paraib{Buy.} This operation is used to purchase inscription tokens from marketplaces. It references the transaction that listed the inscription on the marketplace. The \stress{buy} operation specifies only protocol standards, as the token name (\stress{tick}), amount, and price are declared in the \stress{list} operation.

\paraib{Sell.} It is used for selling inscriptions, this operation specifies the protocols and the order message containing seller and signer details, as well as the amount of tokens being sold and their prices.

\subsection{Comparison with NFTs and ERC-20s}

In this section, we discuss the distinctions between inscriptions and established \stress{\gls{ERC}} standards such as ERC-20~\cite{Token@ERC-20} and ERC-721~\cite{NFT@ERC-721} concerning tokens and \acrfull{NFT}, respectively.

\paraib{Comparison with NFTs.}
In the Ethereum blockchain, \glspl{NFT}~\cite{schaar2022non,wang2021non} are created through smart contracts. Each user receives a unique \stress{token ID}, affirming their ownership of a specific asset. These assets, which can be JPEG files or CryptoPunk images~\cite{schaar2022non}, are stored off-chain on a server or on the \gls{IPFS}~\cite{IPFS}.

In contrast to \glspl{NFT}, inscription-based tokens do not rely on any smart contract and, thus, do not allow upgrades. The link to the asset file is inscribed into the transaction data.
In a blockchain with a maximal native token supply, such as Bitcoin, the amount of inscription is bound by the blockchain network limits. In contrast, \glspl{NFT} based on smart contracts are free from such limitations, theoretically allowing for unlimited minting. Also, each inscription is allocated a position in the blockchain, which creates the opportunity to derive the additional value of the inscription from the location within the block.

\paraib{Comparison with ERC-20.}
The other application of inscriptions is to create ERC-20-style tokens. As inscription-based tokens are not reliant on smart contracts, they are not susceptible to the risk of smart contract upgrades, and their maximal circulating supply is declared once in the \stress{deploy} operation. The minting and transfer of each new inscription-based token can be tracked directly on the blockchain and cannot be altered.

However, due to their lack of smart contract support, inscription-based tokens offer limited functionality. They are primarily utilized for speculative purposes and fall into the category of meme-coins~\cite{wang2023brc}. Based on the inscriptions technology, it is possible to mint a single \gls{NFT}-style token or a group of tokens with a predefined token supply in circulation. Nevertheless, it is the off-chain indexer, managed by the token's creator, responsible for ensuring the data integrity of minted tokens.
Trading of inscription-based tokens predominantly occurs within inscriptions marketplaces since they are not compatible with the ERC-20 standard required by \glspl{DEX}~\cite{xu2023sok}.

\section{Data Collection}\label{sec:dataset}

\begin{table*}[t]
\centering
\scriptsize
\caption{Data set used to analyze inscription events on five blockchains.}
\resizebox{\textwidth}{!}{%
\begin{tabular}{lccrrrr}
\toprule
\setlength{\tabcolsep}{6pt}
\thead{Chain} & \thead{Start date} & \thead{End date} & \thead{\# of Issuers} & \thead{\# of Blocks}      &
\thead{\# of Inscriptions}  \\
\midrule
    Arbitrum & June 17\tsup{th}, 2023 & April 30\tsup{th}, 2024  & \num{118544} & \num{3575299}  & \num{16309035}\\
    Base & July 28\tsup{th}, 2023 & April 30\tsup{th}, 2024  & \num{79573} & \num{780770}  & \num{2020661}\\
    Ethereum & June 14\tsup{th}, 2023 & April 30\tsup{th}, 2024  & \num{245008} & \num{930824}  & \num{6493580}\\
    Optimism & June 18\tsup{th}, 2023 & April 30\tsup{th}, 2024  & \num{49112} & \num{588053}  & \num{1475663}\\
    ZKsync Era & June 18\tsup{th}, 2023 & April 30\tsup{th}, 2024  & \num{481687} & \num{2809054}   & \num{17161306}\\
\bottomrule
\end{tabular}}
\label{tab:dataset}
\end{table*}

We analyze the inscriptions data recorded on Ethereum and its major \gls{EVM}-compatible rollups: ZKsync Era, Arbitrum, Optimism, and Base. From these \glspl{L2}, ZKsync Era is a \gls{ZK} rollup, whereas other chains are optimistic rollups.

We then collected blockchain data from various sources and made them and our scripts available in a public repository~\cite{Messias-DataSet-Code-2025}. For ZKsync, we use data obtained from their official archive node and a Web3-compatible API for ZKsync Era~\cite{ZKsync-Node}. For Arbitrum, Base, Ethereum, and Optimism, we use data sources from Nansen BigQuery~\cite{NansenQuery}. These data sources enable us to collect comprehensive data containing all information recorded on the evaluated chains. This includes data about blocks, transactions, and events (or logs triggered during the execution of smart contracts) specifically related to inscriptions. Table~\ref{tab:dataset} provides an overview of the specifics of our dataset.

Our focus is on blockchain data associated with inscriptions. While a significant portion of inscriptions involves the issuance of self-transfer transactions (i.e., where issuers initiate a transfer to their own address), we adopt a broader perspective. This involves considering instances where inscriptions are added to the chain, irrespective of whether through a self-transfer or not. This is important to capture the total fees users spent on inscriptions.

To identify these inscriptions, we specifically search for transactions with input call data starting with ``\verb|0x646174613a|'', representing ``\verb|data:|'' in ASCII. This criterion allows us to recognize inscriptions within the blockchain. The blockchains with the most number of transaction inscriptions in our dataset were ZKsync Era and Arbitrum. In total, we found \num{17054466} transactions containing inscriptions in our ZKsync Era data set issued by~\num{481687} addresses and added to~\num{2809054} blocks. Similarly, we identified \num{16309035} transactions containing inscriptions in Arbitrum, issued by \num{118544} addresses and added to \num{3575299} blocks.
The average number of inscriptions per block are~\num{6.07} and \num{4.56} for ZKsync Era and Arbitrum, respectively. We provide similar statistics for Base, Ethereum, and Optimism chains in Table~\ref{tab:dataset}. This dataset forms the basis for our empirical analysis of inscriptions.

\begin{table}[t]
\centering
\caption{Inscription transactions per issuers.}
\resizebox{0.7\textwidth}{!}{%
\begin{tabular}{lrrrrrr}
\toprule\small
\thead{Protocol} & \thead{\# of Issuers} & \thead{Mean} & \thead{Std.} & \thead{Median} & \thead{Min} & \thead{Max} \\
\midrule
Arbitrum & \num{118544}&\num{137.57} & \num{676.76} & \num{6} & \num{1} & \num{38050} \\
Base & \num{79573}&\num{25.39} & \num{154.91} & \num{3} & \num{1} & \num{19674}  \\
Ethereum & \num{245008}&\num{26.50} & \num{245.60} & \num{3} & \num{1} & \num{67713}  \\
Optimism & \num{49112}&\num{30.04} & \num{168.05} & \num{3} & \num{1}  & \num{19612}  \\
ZKsync Era & \num{481687}&\num{35.62} & \num{245.35} & \num{3} & \num{1} & \num{40770} \\
\bottomrule
\end{tabular}}
\label{tab:issuers}
\end{table}

The distribution of inscription transactions per address is highly skewed towards a few participants, as shown in Table~\ref{tab:issuers}. Most inscription issuers initiated at least three transactions related to inscriptions. Arbitrum exhibited the highest median, with at least six transactions per issuer. Table~\ref{tab:issuers} also highlights that a small number of issuers were responsible for a significant fraction of the total inscriptions. For instance, a single issuer accounted for \num{67713} transactions to claim their inscriptions, followed by ZKsync Era with \num{40770} transactions. These are likely bots accounts.

\section{Empirical Analysis}
\label{sec:characterization}

In this section, we present our empirical analysis of the inscriptions recorded on Ethereum and its major \gls{EVM}-compatible rollups: Arbitrum, Base, Optimism, and ZKsync Era. We characterize the inscription protocols, operations, tokens, and their trading dynamics. Then, we investigate the impact of sudden surges in inscription transactions on the blockchains' performance.

\begin{figure*}[t]
\centering
\begin{subfigure}{1\onecolgrid}
\centering
        \includegraphics[width=\textwidth]{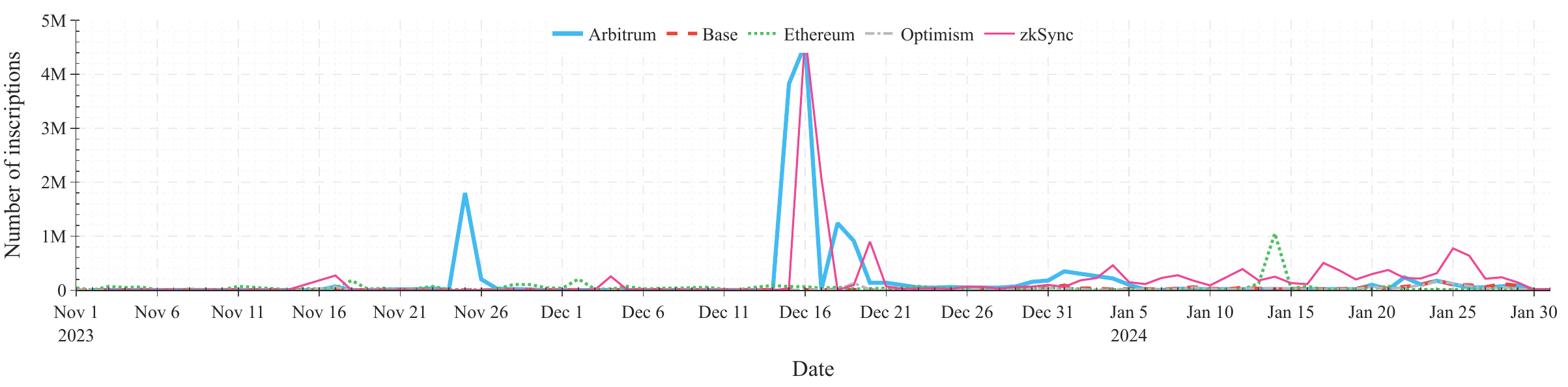}
        \caption{Absolute number of inscriptions.}
        \label{fig:tx-inscription-dist-transactions}
\end{subfigure}

\begin{subfigure}{1\onecolgrid}
\centering
        \includegraphics[width=\textwidth]{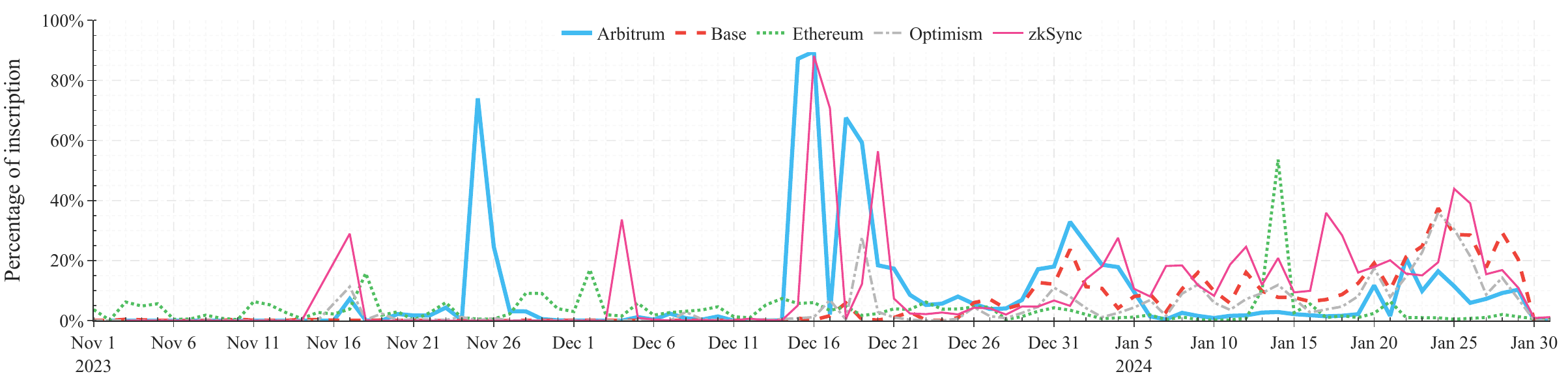}
        \caption{Percentage of inscriptions in comparison to the overall number of transactions.}
        \label{fig:tx-inscription-dist-fraction}
\end{subfigure}
\caption{Daily distribution of inscription transactions in our data set: (a) absolute number of inscription transactions; and (b) percentage of inscription transactions compared to the total number of transactions.}
\label{fig:tx-inscription-dist}
\end{figure*}

\subsection{Overall Transactions}
\label{subsec:overall-txs}

Our analysis shows multiple spikes in inscription transaction counts occurred on Arbitrum (November~25\tsup{th}, December~15\tsup{th},~16\tsup{th}, and~18\tsup{th}) and ZKsync Era (November~17\tsup{th},~16\tsup{th},~17\tsup{th}, and December 21\tsup{st}) as shown in Fig.~\ref{fig:tx-inscription-dist-transactions}.
During peak periods, inscriptions dominated daily transactions, comprising nearly 90\% on ZKsync Era and Arbitrum (December~16\tsup{th},~2023), over~53\% on Ethereum (January~14\tsup{th},~2023), and almost~35\% on Optimism and Base (January 24\tsup{th},~2024). Fig.~\ref{fig:tx-inscription-dist-fraction} illustrates those peaks. Due to this high activity of inscriptions on-chain, some rollups were not prepared to handle the surge in transactions, leading to downtimes and delays in the processing of transactions~\cite{Tom-Arbitrum@Cointelegraph,Haig@TheDefiant}. Next, we investigate the protocols and operations that contribute to these observed spikes.

\subsection{Inscriptions Characterization}
\label{subsec:zksync}

\begin{figure}[t]
  \centering
        \includegraphics[width=\onecolgrid]{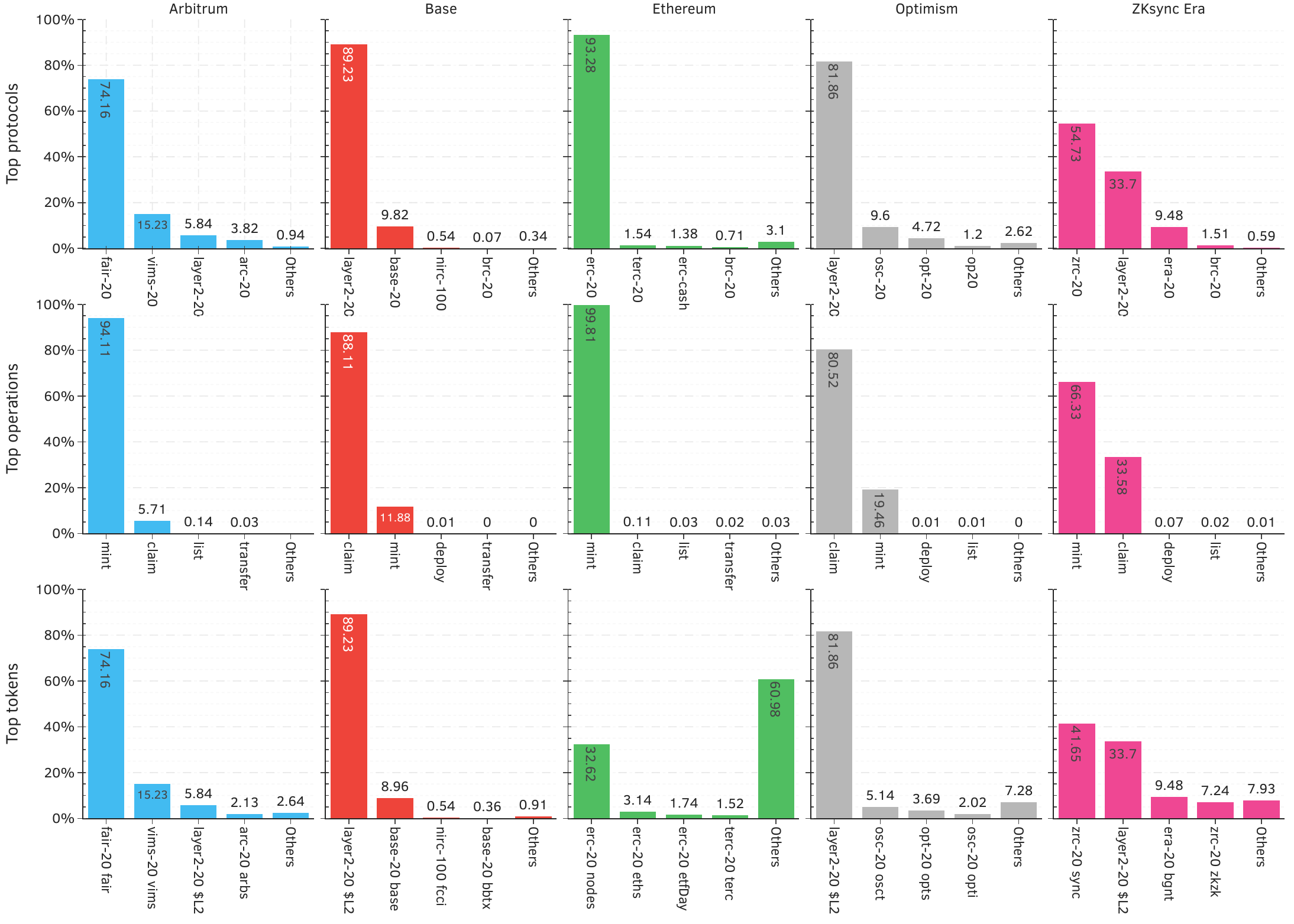} 
    \caption{Percentage breakdown of the top~3 inscriptions attributes~---~protocol, operations, and tokens~---~for the chains analyzed in our study: Arbitrum, Base, Ethereum, Optimism and ZKsync Era.}
    \label{fig:all_analysis}
\end{figure}

Our analysis indicates that inscription tokens are uniquely unified by a protocol-tick pair. Therefore, we aggregated data using this pair to identify the top inscriptions tokens. Fig.~\ref{fig:all_analysis} illustrates the breakdown of inscription transactions based on protocol, operation, and the leading token. In the following, we provide our observations for each of these categories.

\paraib{Protocols.}
Our initial observation reveals, in (Fig.~\ref{fig:all_analysis} row \stress{Top Protocols}), that each blockchain contains a particular leading inscription protocol, similar to \stress{BRC-20} on Bitcoin~\cite{yu2023bridging,wang2023understanding}, constituting between 54\% and 93\% of transactions. In addition, blockchain-specific standards such as \stress{FAIR-20} for Arbitrum,  \stress{ERC-20} for Ethereum, and \stress{ZRC-20} for ZKsync Era appear the most.
The \stress{LAYER2-20} protocol is present across all studied \glspl{L2}, being the dominant inscription protocol on Base and Optimism, and the second largest on ZKsync Era.

\paraib{Operations.}
Recall from \S\ref{sec:background_inscriptions} that both \stress{claim} and \stress{mint} operation types work similarly. Thus, if evaluated together they account for approximately~99\% of all inscription transactions in all chains we studied.
From Fig.~\ref{fig:all_analysis} row \stress{Top Operations}, a small number of transactions are attributed to the \stress{list} operation, used for listing inscriptions for sale in an inscription marketplace, and the \stress{deploy} operation, which declares the token on the chain along with its token supply. The low percentage of \stress{transfer}, \stress{sell}, and \stress{buy} transactions suggests that inscription tokens are not yet actively traded on \gls{EVM} chains.

\paraib{Tokens.}
From our analysis (Fig.~\ref{fig:all_analysis}, row \textit{Top Tokens}), we observe that various protocols deploy and mint tokens with the same tick symbols. To account for this, we identify each token using a protocol-tick pair. Our findings indicate that Arbitrum, Base, and Optimism each have a dominant token, comprising between~74\% and~89\% of inscription transactions on their respective platforms. In contrast, the token distribution on Ethereum is more diverse, with the \textit{ERC-20 NODES} token contributing approximately 32\% of the transactions, while all other tokens individually represent less than~3.14\%. On ZKsync Era, the top four tokens account for over~92\% of inscription transactions, also indicating token distribution concentration when compared to other chains. Notably, the token \stress{LAYER2-20 \$L2} is present across all rollups and is also the dominant token on both Base and Optimism.

\subsection{Inscription Trading}
\begin{figure}[t]
    \centering
\begin{subfigure}{\twocolgrid}
       \centering
        \includegraphics[width=\twocolgrid]{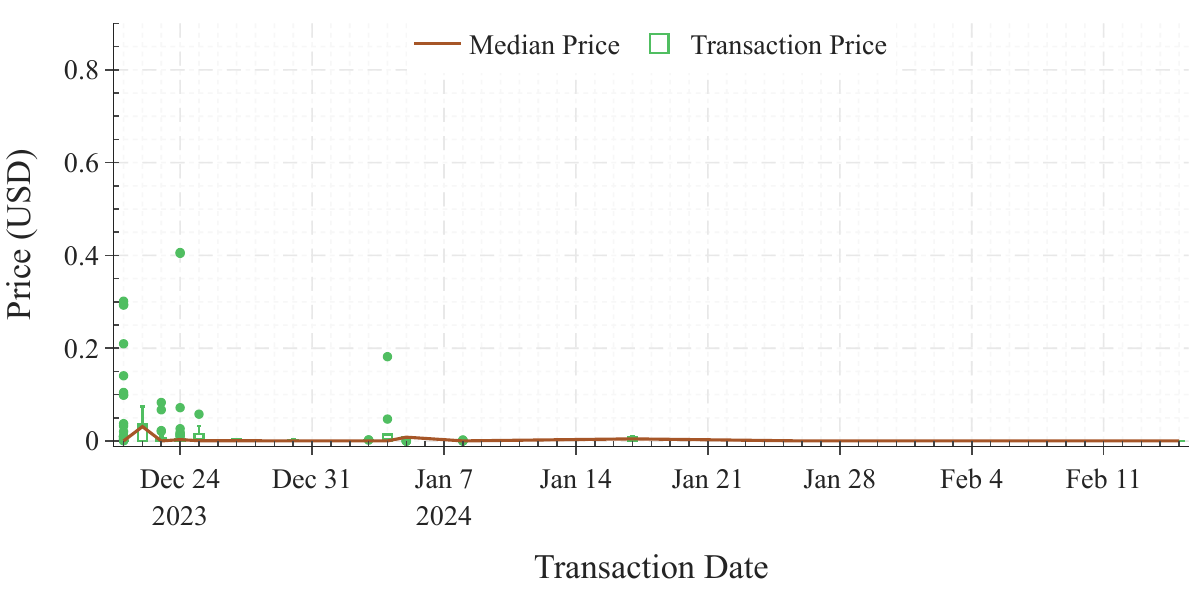}
        \caption{\stress{Transfer transactions of zrc-20 sync}.}
        \label{fig:transfersync}
\end{subfigure}
\begin{subfigure}{\twocolgrid}
        \centering
        \includegraphics[width=\twocolgrid]{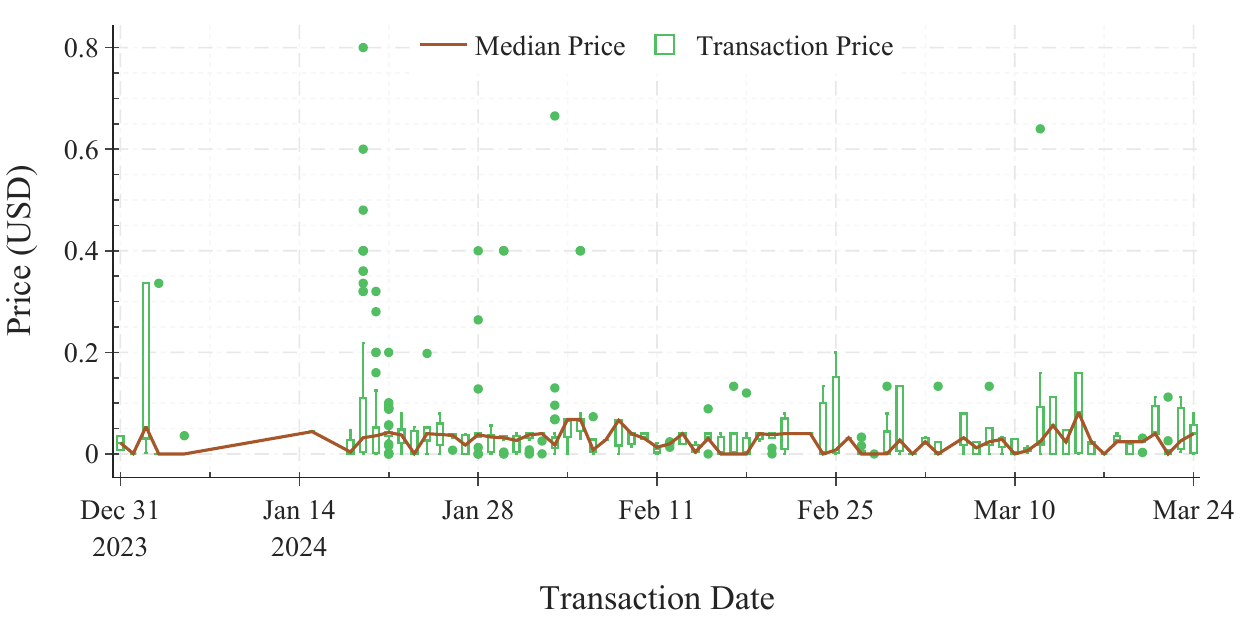}
        \caption{Buy transactions of \stress{era-20 bgnt}.}
        \label{fig:buybgnt}
\end{subfigure}
\caption{Price per token at every transaction on ZKsync Era considering \stress{zrc-20 sync} and \stress{era-20 bgnt} tokens.}
\label{fig:transaction_price}
\end{figure}

We analyze the trading behavior of two inscription tokens within the ZKsync Era: \stress{zrc-20 sync} (the predominant token) and \stress{era-20 bgnt} (a smaller but noteworthy token due to its completed minting process). Our analysis, presented in Fig.~\ref{fig:transaction_price}, shows their respective trading activities. Notably, \stress{era-20 bgnt} has a higher level of trading activity compared to \stress{zrc-20 sync}. Specifically, we observed~\num{477} transfer transactions for \stress{zrc-20 sync}, along with the corresponding token median prices as shown in Fig.~\ref{fig:transfersync}, and~\num{1148} buy transactions for \stress{era-20 bgnt}, with corresponding median prices depicted in Fig.~\ref{fig:buybgnt}

Our analysis shows that the price of \emph{zrc-20 sync} oscillated between~\$0.10 and~\$0.4, while \emph{era-20 bgnt} ranged from~\$0.10 to~\$0.8. The trading of the \emph{era-20 bgnt} token continued from its minting throughout the entire studied period. As the trades occur in batches of~4 or~5 tokens, they provide minimal compensation to the seller to cover the gas fees necessary to mint and list the inscriptions. However, trading inscriptions at such low prices is feasible due to the low gas fees on rollups, which ranged at ZKsync Era from~\$0.05 to~\$0.25 (see Fig.~\ref{fig:gasfees}). 

\begin{table}[t]
\centering
\caption{Inscription operations cost in GWei for major tokens.}
\resizebox{\textwidth}{!}{%
\begin{tabular}{ccccccccc}
\toprule
\thead{Chain} & \thead{Token} & \thead{Operation} & \thead{Total (in ETH)} & \thead{Mean} & \thead{Std.} & \thead{Median} & \thead{Min} & \thead{Max} \\
\midrule
Arbitrum & \stress{fair-20 fair} & \stress{mint} & \num{0.77} & \num{74.10} & \num{305.72} & \num{60.87} & \num{0} & \num{488849.89} \\
Arbitrum & \stress{layer2-20 \$L2} & \stress{claim} & \num{0.03} & \num{40.12} & \num{18.83} & \num{34.77} & \num{0.57} & \num{500.85} \\
Base & \stress{layer2-20 \$L2} & \stress{claim} & \num{0.0016}& \num{1.03} & \num{8.27} & \num{0.02} & \num{0} & \num{2199.20} \\
Ethereum & \stress{erc-20 nodes} & \stress{mint} & \num{0.19} & \num{176.70} & \num{7015.97} & \num{99.32} & \num{0.01} & \num{1351757.78} \\
Optimism & \stress{layer2-20 \$L2} & \stress{claim} & \num{0.0009} & \num{0.78} & \num{3.04} & \num{0.11} & \num{0} & \num{1617.84} \\
ZKsync Era & \stress{zrc-20 sync} & \stress{mint} & \num{88.86} & \num{12684.59} & \num{3795.42} & \num{12748.95} & \num{0} & \num{1097718.16} \\
ZKsync Era & \stress{era-20 bgnt} & \stress{mint} & \num{11.31} & \num{7114.49} & \num{2525.91} & \num{6292.83} & \num{0.32} & \num{244836.15} \\
ZKsync Era & \stress{era-20 bgnt} & \stress{list} & \num{0.015} & \num{4839.49} & \num{9528.53} & \num{272.57} & \num{0} & \num{60807.7} \\
ZKsync Era & \stress{layer2-20 \$L2} & \stress{claim} & \num{0.017} & \num{29.63} & \num{6.90} & \num{28.38} & \num{0.42} & \num{2001.14} \\
\bottomrule
\end{tabular}}
\label{tab:fees_per_token}
\end{table}

Next, we analyze how spikes in inscription transactions influence gas fees across the network. Table \ref{tab:fees_per_token} presents a breakdown of fees incurred during inscription operations. Recall that, in the first transaction, \stress{deploy}, the maximum number of tokens that can be minted or claimed in a single transaction is specified.
Then, we calculate the average fee per single token for each operation. We found that the mint operation is more efficient than \stress{claim}, although both have a similar functionality.
The interesting comparison provides \stress{claim} operation for \stress{layer2-20 \$L2} token, as it is operating across all rollups. Its mining (claiming) operation had a lower median fee on Optimism and Base, followed by ZKsync Era and Arbitrum.

\subsection{Impact on Gas Fees}
\label{subsec:ImpactGas}

\begin{figure*}[t]
\centering
\begin{subfigure}{1\onecolgrid}
\centering
        \includegraphics[width=\textwidth]{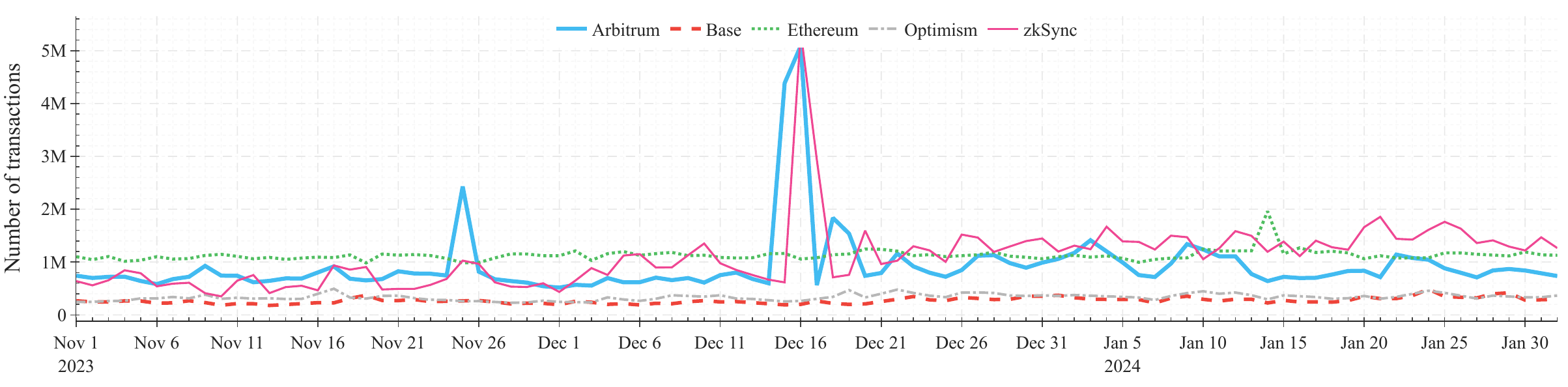}
        \caption{Transaction count.}
        \label{fig:gasfees-tx-count}
\end{subfigure}
\begin{subfigure}{1\onecolgrid}
\centering
        \includegraphics[width=\textwidth]{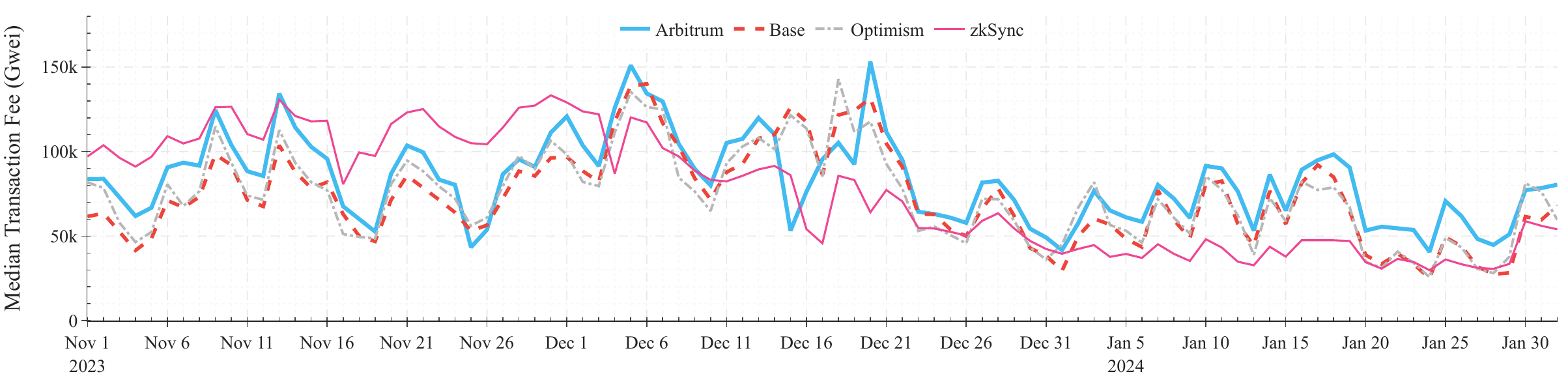}
        \caption{Median transaction cost.}
        \label{fig:gasfees-tx-median-cost}
\end{subfigure}
\caption{Daily transaction count and median transaction cost in USD paid by the user: (a) transactions count; and (b) median transaction cost.}
\label{fig:gasfees}
\end{figure*}

\begin{figure*}[t]
\centering
\begin{subfigure}{\twocolgrid}
\centering
        \includegraphics[width=\twocolgrid]{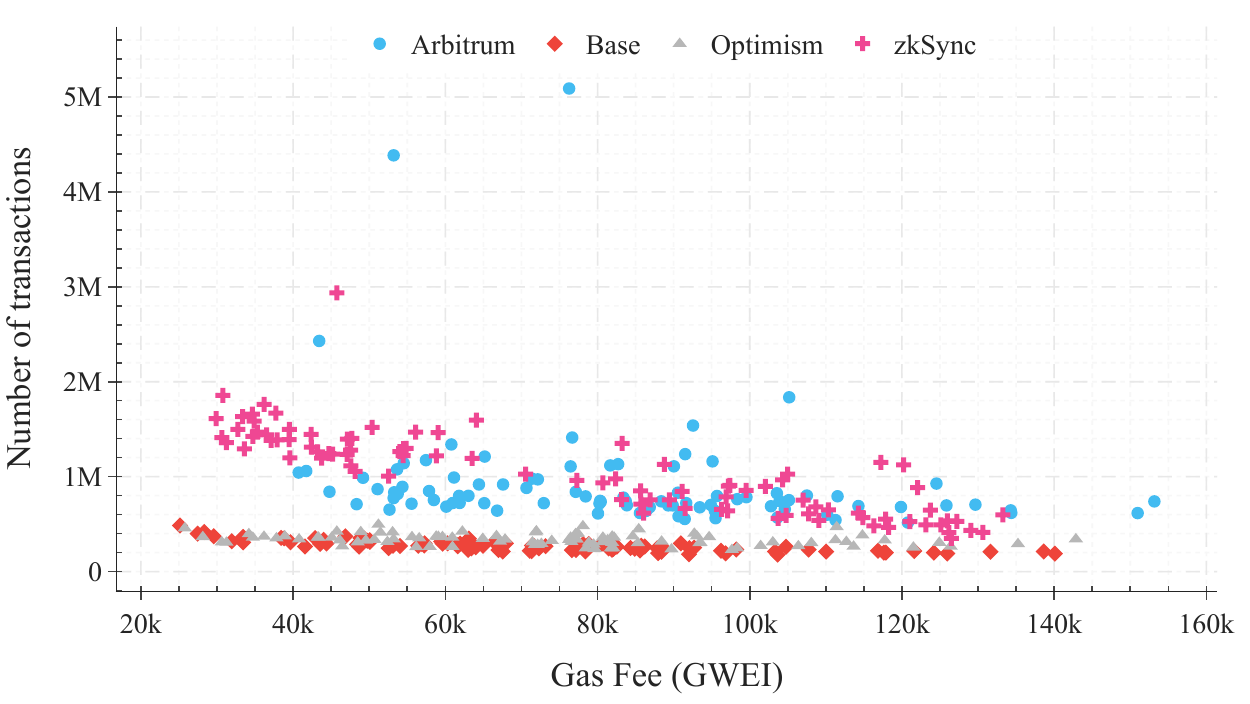}
        \caption{L2 rollups.}
        \label{fig:rollup-regression}
\end{subfigure}
\begin{subfigure}{\twocolgrid}
\centering
        \includegraphics[width=\twocolgrid]{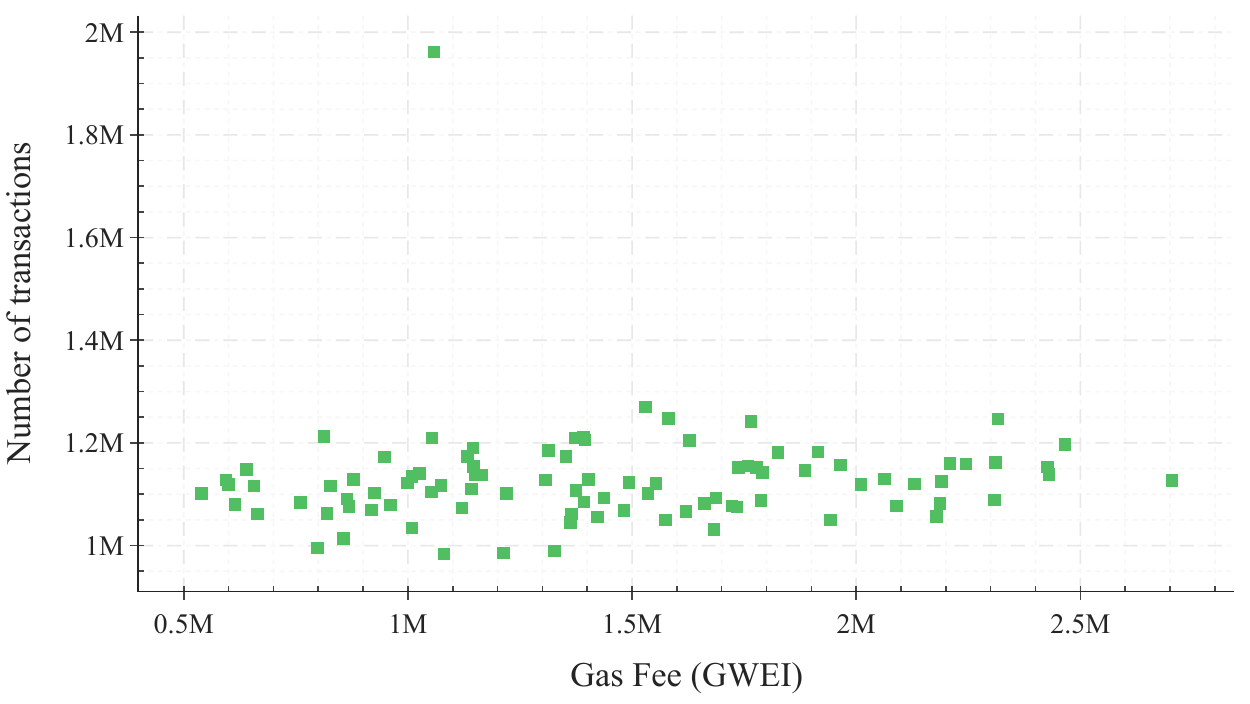}
        \caption{Ethereum mainnet (\gls{L1}).}
        \label{fig:Ethereum-regression}
\end{subfigure}
\caption{Distribution of gas fees in GWei relative to transaction count: (a) L2 rollups; and (b) Ethereum mainnet.}
\label{fig:gas-regression}
\end{figure*}

This section analyzes the impact of daily transaction volumes on the median gas fee per transaction. Unlike \gls{L1} blockchains, rollups are expected to offer lower gas fees with increasing transaction counts due to their scaling mechanism. As more \gls{L2} transactions can be compressed into a single batch stored in the L1 network, gas fees tend to decrease. In particular, spikes in daily transaction counts on specific days resulted in a decrease in median gas fees per transaction for other users, as shown in Fig.~\ref{fig:gasfees}. Fees on \gls{L2} rollups are linked to compression and are also affected by the fees of \gls{L1} Ethereum. 

Further analysis of the relationship between transaction count and median gas fee is presented in Fig.~\ref{fig:gas-regression}. Generally, gas fees with more transactions on rollups either remain unchanged (Base, Optimism) or decrease (Arbitrum, ZKsync). Conversely, on Ethereum, gas fees tend to increase with higher transaction counts. Additionally, ZKsync demonstrates higher efficiency in reducing gas fees compared to Arbitrum, attributable to its \gls{ZK} rollup design, while Arbitrum, like Base and Optimism, is an optimistic rollup. 
Thus, unlike L1 blockchains, gas costs on \gls{ZK} rollups decrease with a larger transaction volume, as observed on November~17, December~5, December~17, and December~20 on ZKsync Era.

\section{Discussion}
\label{sec:discussion}

In this section, we discuss various aspects of inscriptions, starting with the impact of blobs on inscription ownership. We then explore whether inscriptions are beneficial or detrimental to the blockchain ecosystem. Finally, we address the potential for \stress{pump-and-dump} schemes~\cite{li2021cryptocurrency,xu2019anatomy} associated with inscriptions.

\paraib{Impact of blobs on inscription ownership.}
As inscriptions are appended to transactions as arbitrary data within the input call data, there is no assurance of perpetual availability of this data in the future. Blobs, initially designed to increase the efficiency of Ethereum rollups~\cite{2024EthereumRoadmap}, were used to issue new inscription directly in Ethereum blockchains, called \stress{BlobScriptions}~\cite{2024DuneEVMInsctiptions}. Yet, blob data is intended to be stored by Ethereum nodes for~18 days, during which rollup operators or other parties verify their transactions. After this period, BlobScriptions disappear from the Ethereum blockchain and are only stored (and available) in the indexer of the creator protocol, outside the blockchain. Similarly, rollups are not required to store (outside of blobs) the transaction input data used by rollup-based inscriptions, raising questions about the ownership of inscriptions.

\paraib{Inscriptions: good or bad?}
A report by Binance on BRC-20 highlights that the unexpected surge in transactions resulted in an increase in transaction fees for many blockchains~\cite{BRC-20@Binance}. The report suggests that such an increase in fees can be seen as a natural progression in blockchain adoption.
Therefore, attribution of this spike solely to BRC-20s is considered irrelevant. More concerning, however, is that other protocols, such as Arbitrum, experienced downtime of approximately 78 minutes~\cite{Tom-Arbitrum@Cointelegraph}, while ZKsync Era saw an increase in \gls{TPS}, with nearly 96\% of all transactions being related to inscriptions over the course of several hours. Thus, causing users to worry about delayed transaction inclusion. This sparked a broader debate within the blockchain community about whether ordinals and inscriptions truly benefit users.

\paraib{Challenges and opportunities for trading inscription.}
Inscriptions are not compatible with ERC-20 tokens. Therefore cannot be traded on traditional decentralized exchanges based on \glspl{AMM} such as Uniswap~\cite{adams2021uniswap}, which is the base for trading tokens with low market capitalization. Today, inscriptions are listed at \gls{NFT}-marketplaces in a limited-order book type, reducing their liquidity~\cite{BigInt@Marketplace,Inscribe@Marketplace,ZKSMarket@Marketplace}. Also, for many inscription-based meme-coins, their minting process may still be ongoing, and new tokens could continue to be minted until the token limit specified in the \stress{deploy} operation is reached. Furthermore, wrapping inscription-based meme coins into ERC-20 tokens could allow these inscriptions to be listed on \glspl{DEX}, attracting meme-coin traders who are not familiar with \gls{NFT} marketplaces. We see this as the next evolution of inscriptions if they are intended to stay.

\paraib{Hope or Hype?}
During minting periods, platforms utilized various channels to engage users, including communities, influencers, and specific chains that amplified them. Consequently, numerous users issued inscriptions across different platforms~\cite{BigInt@Marketplace,Inscribe@Marketplace,ZKSMarket@Marketplace}.
Understanding user expectations, often influenced by \gls{FOMO}, regarding inscriptions is crucial. For instance, \stress{are inscriptions more about hope or hype?} This question relates to whether inscriptions are a novel concept or just a passing trend. In the case of Bitcoin, there was a significant surge in interest and user participation during the BRC-20 movement, followed by a subsequent decline in enthusiasm over the following months~\cite{wang2023brc,wang2023understanding}. Our work shows similar results for \gls{EVM}-compatible chains: \stress{inscriptions peak initially but witness a decline in user interest over time}, as shown in Section~\ref{subsec:overall-txs}.
The evolution of inscriptions fuels discussions on innovative features and integration with tokens and \glspl{NFT}. Despite potential profitability, marketplace platforms encounter challenges in translating inscriptions into substantial profits for users, particularly across different chains. Due to the nature of inscriptions, which rely on off-chain applications, there is currently no mechanism to prevent users from purchasing previously sold inscriptions. Additionally, users can utilize software programs to automatically claim the majority of inscriptions during the minting or claiming period. Hence, \stress{should inscriptions be distributed directly to wallets, similar to traditional airdrop methods~\cite{messias2023airdrops}?}

Addressing these questions could potentially improve inscriptions, transforming them into a promising addition to the blockchain ecosystem.

\section{Related Work}\label{sec:related-work}
In this section, we review the existing literature and related works pertinent to our study on inscriptions within \gls{EVM}-compatible chains. 

\paraib{Social phenomena on financial markets.}
The impact of social phenomena on financial markets has received some research attention recently. In traditional finance, the GameStop short-squeeze event from 2021 highlighted how online social movements can lead to significant shifts in trading behavior and even impact established financial institutions~\cite{zheng2021game}.  Later, various authors analyzed this unprecedented event in more detail. Vasileiou et al.~\cite{vasileiou2021explaining} found a causal link between the stock performance in early 2021 and the Google Trend Index for the stock, thus showing the impact of online interest in the stock in its price performance. Following this, Anand and Pathak~\cite{anand2022role} and Zheng et al.~\cite{zheng2021game} focussed their analysis on the activity of the subreddit \stress{r/wallstreetbets} (where the majority of the social activity related to GameStop was occurring during that time). Their findings further corroborate the link between stock performance and the subreddit's social media activity and sentiment.

\paraib{Social phenomena on decentralized finance.}
Even though the introduction of ordinals and inscriptions is fairly recent, there has been some work towards understanding these phenomena. Wang et al.~\cite{wang2023brc,wang2023understanding} conducted a comprehensive analysis of BRC-20 tokens within the Bitcoin network. Their investigation revealed a notable surge in interest and user participation during the BRC-20 movement. The study involved a comparison between these BRCs and the widely-used Ethereum ERC-20 tokens, employing metrics such as average price return, volatility, and other relevant factors. The study highlighted a significant influx of users into the BRC market within a month, coupled with a subsequent decline in enthusiasm over the following months. In essence, users displayed a tendency to lose interest after the initial surge. In the broader context, the research emphasized that, despite the hype, BRC-based tokens constituted only a modest fraction of the overall market size when juxtaposed with ERC-like tokens on the Ethereum platform. Our paper also analyses the broader market dynamics of inscriptions during their boom, however, we expand the study to the top \gls{EVM} chains instead of focussing on Bitcoin ordinals.

Li et al.~\cite{li_bitcoin_2024} also looked at BRC-20 tokens within the Bitcoin network. However, they perform a comparative analysis between Bitcoin inscriptions and NFTs focussed on technical implementation details, function, and use cases. They additionally provide an in-depth historical overview of Bitcoin-related NFTs. While our paper also involves a characterization of inscriptions and their functions, our focus remains on EVM Layer 2s rather than Bitcoin.

Finally, Bertucci~\cite{bertucci2023bitcoin} analyzed the impact of ordinals on transaction fees in Bitcoin. They found that ordinal inscriptions typically have lower fee rates than regular transactions, which indicates careful fee-setting by ordinal users. They also concluded that the rise of ordinal transactions has increased non-inscription transaction fee rates and total block fees, thus increasing miners' revenue. We do a similar fee investigation in our paper. Yet, we look at inscriptions in \gls{EVM} chains instead of Bitcoin.

\paraib{Novelty of our approach.}
To the best of our knowledge, this is the first work to conduct an in-depth analysis of inscriptions in multi \gls{EVM}-compatible chains and the user dynamics that arose from the inscription boom observed in the end of 2023 and early 2024. While Bitcoin originally set the stage for this user trend~\cite{wang2023brc,wang2023understanding}, \gls{EVM} chains have witnessed substantial adoption as well. The absence of analysis in this domain introduces a notable gap in the existing research landscape, which we address in this paper.

\section{Conclusion}\label{sec:conclusion}

The inscriptions boom between December~2023 and January~2024 was the first real-life stress test for various \gls{EVM}-compatible blockchains. Our study of Ethereum, Arbitrum, Base, Optimism, and ZKsync Era during this period revealed stark differences between \gls{L1}, \gls{L2} blockchains, and among rollups. 

First, we observed that, on specific peak days, transactions related to inscriptions accounted for nearly 90\% on Arbitrum and ZKsync Era, and over 53\% on Ethereum, and almost 35\% on Optimism and Base. The surge in on-chain inscriptions overwhelmed some rollups, causing their instability due to the increased transaction volume. This resulted in downtimes and delays in transaction processing.

We further found that 99\% of these inscription transactions across all examined blockchains were driven by the claim or minting of inscription-based meme-coins, followed by a modest amount of trading activity, mostly associated with listing operations on marketplaces. All chains show a dominant token that represents more than 74\% of all inscription transactions. 

Looking at the impact of inscriptions on gas fees, we concluded that gas fees on rollups either remained stable or decreased during the transaction volume peak. Particularly, the median transaction fee remained the same on Base and Optimism, and decreased on Arbitrum and ZKsync Era. On the other hand, Ethereum experienced a rise in fees, which is not surprising. Unlike \gls{L1} blockchains, rollups are designed to offer lower gas fees as transaction counts increase, due to their ability to compress multiple \gls{L2} transactions into a single batch on the \gls{L1} network.

Moreover, the future of inscriptions as a stable market for token trading in \gls{EVM}-chains is still not cemented. For instance, \gls{DA} and \gls{UI} issues can complicate interactions with inscriptions, highlighting how nascent the market still is. It is nevertheless an interesting trend and a lot of innovation can still occur to turn the space into a more mature one.

Finally, to promote scientific reproducibility and facilitate further research, we made our data set and scripts publicly available~\cite{Messias-DataSet-Code-2025}.


\bibliographystyle{splncs04}
\bibliography{references}

\appendix

\section{Glossary}
\label{sec:glossary}
Following is a list of important notations used in this paper.
\setglossarystyle{alttree}
\glssetwidest{AAAA}
\printnoidxglossary[type={acronym}]

\end{document}